\def\complexNumbers{\mathbb{C}}

\def\integers{\mathbb{Z}}
\def\integersPositive{\mathbb{Z}^{+}}
\def\integersNonnegative{\mathbb{Z}_{0}^{+}}

\def\constante{{\rm e}}
\def\constantj{{\rm j}}
\def\constanti{{\rm i}}
\def\constantMinusi{{\rm j}}
\def\constantOne{{\rm +}}
\def\constantMinusOne{{\rm -}}

\def\separationVar{m}

\def\upsampleVarA{k}
\def\upsampleVarB{l}

\def\lagForCorrelation{k}

\def\indexIterationANF{l}

\def\orderMonomial[#1]{k_{#1}}
\def\coeffientsANF[#1]{c_{#1}}
\def\polyVariable{z}

\def\modulationSymbolF[#1]{m_{#1}}
\def\cardinalitySetOfOperators[#1]{{H}_{#1}}

\def\monomial[#1]{x_{#1}}

\def\lengthGaGb{L_{\rm s}}
\def\lengthGcGd{M}

\def\scaleAexp[#1]{a_{#1}}
\def\scaleBexp[#1]{b_{#1}}
\def\scaleEexp[#1]{e_{#1}}
\def\angleexp[#1]{c_{#1}}

\def\separationGolay[#1]{d_{#1}}

\def\eleGa[#1]{{a}_{#1}}
\def\eleGb[#1]{{b}_{#1}}

\def\apac[#1][#2]{\rho_{#1}(#2)}
\def\apacPositive[#1][#2]{\rho^{+}_{#1}(#2)}
\def\binaryAsignment[#1][#2]{b_{#1}^{(#2)}}

\def\eleSeqf[#1]{{f}_{#1}}
\def\eleSeqg[#1]{{g}_{#1}}
\def\eleSeqcf[#1]{{c}_{f,#1}}
\def\eleSeqcg[#1]{{c}_{g,#1}}

\def\scaleA[#1]{\alpha_{#1}}
\def\scaleB[#1]{\beta_{#1}}
\def\angleGolay[#1]{\omega_{#1}}
\def\permutationShift[#1]{{\psi_{#1}}}
\def\permutationMono[#1]{{\pi_{#1}}}

\def\symbolDuration{T_{\rm s}}
\def\timeVar{t}

\def\funczArbitrary[#1]{z(#1)}
\def\funcaArbitrary[#1]{a(#1)}
\def\funcbArbitrary[#1]{b(#1)}
\def\coefficientArbitrary[#1]{k_{#1}}

\def\vecArrangement[#1]{\textbf{b}_{#1}}

\def\seqGa{\textit{\textbf{a}}}
\def\seqGb{\textit{\textbf{b}}}
\def\seqGaIt[#1]{\textit{\textbf{a}}^{(#1)}}
\def\seqGbIt[#1]{\textit{\textbf{b}}^{(#1)}}
\def\seqGc{\textit{\textbf{c}}}
\def\seqGd{\textit{\textbf{d}}}
\def\seqGf{\textit{\textbf{f}}}
\def\seqGg{\textit{\textbf{g}}}
\def\seqGfdot[#1]{\bar{\textit{\textbf{f}}}_{#1}}
\def\seqGgdot[#1]{\bar{\textit{\textbf{g}}}_{#1}}

\def\seqSub[#1]{\textit{\textbf{h}}_{#1}}

\def\seqFirstOrderMonomial[#1]{\textit{\textbf{m}}_{#1}}

\def\seqToBeModulated[#1]{\textit{\textbf{s}}_{#1}}

\def\flipConjugate[#1]{{{\tilde{#1}}}}
\def\expectationOperator[#1]{{\mathbb{E}}[#1]}
\def\operator[#1][#2]{\mathcal{O}_{#1}^{(#2)}}
\def\operatordot[#1][#2]{\bar{\mathcal{O}}_{#1}^{(#2)}}

\def\compositeOperatorF[#1][#2]{{F}_{#1}{(#2)}}
\def\compositeOperatorG[#1][#2]{{G}_{#1}{(#2)}}
\def\compositeOperatorFdot[#1][#2]{\bar{F}_{#1}{(#2)}}
\def\compositeOperatorGdot[#1][#2]{\bar{G}_{#1}{(#2)}}

\def\setOfOperators[#1]{{\mathfrak{J}}_{#1}}

\def\operatorBinary[#1][#2]{O_{#1}^{(#2)}}
\def\operatorSign[#1][#2]{{\rm S}_{#1}^{(#2)}}
\def\operatorScaleA[#1][#2]{{\rm{A}}_{#1}^{(#2)}}
\def\operatorScaleB[#1][#2]{{\rm{B}}_{#1}^{(#2)}}
\def\operatorAngle[#1][#2]{\Omega_{#1}^{(#2)}}
\def\operatorSeparation[#1][#2]{\Delta_{#1}^{(#2)}}
\def\operatorOrderA[#1][#2]{\dot{\rm O}_{#1}^{(#2)}}
\def\operatorOrderB[#1][#2]{\ddot{\rm O}_{#1}^{(#2)}}

\def\upsampleOp[#1][#2]{{\uparrow_{#1}\{#2\}}}

\def\functionf[#1]{p^{(#1)}}
\def\functiong[#1]{q^{(#1)}}

\def\functionfdot[#1]{\bar{p}_{\indexIterationANF}^{(#1)}}
\def\functiongdot[#1]{\bar{q}_{\indexIterationANF}^{(#1)}}

\def\funcGfForANF[#1]{f_{#1}}
\def\funcGgForANF[#1]{g_{#1}}
\def\polySeq[#1][#2]{p_{#1}(#2)}

\newcommand\mydots{\hbox to 1em{.\hss.\hss.}}

\documentclass[journal]{IEEEtran}

\usepackage{multicol}
\usepackage{lipsum}
\usepackage{mathtools}
\usepackage{amsthm}
\usepackage{acronym}
\usepackage{stfloats}
\usepackage{xcolor}
\usepackage{amsfonts}
\usepackage{acronym}
\usepackage{cite}
\usepackage{bm}
\usepackage{amsmath,amssymb}
\usepackage{graphicx}
\usepackage[english]{babel}
\usepackage[caption=false,font=footnotesize]{subfig}
\usepackage[geometry]{ifsym}
\usepackage{array}
\usepackage[utf8]{inputenc}
\usepackage[T1]{fontenc}

\usepackage{cuted}
\makeatletter
\newif\ifAC@uppercase@first%
\def\Aclp#1{\AC@uppercase@firsttrue\aclp{#1}\AC@uppercase@firstfalse}%
\def\AC@aclp#1{%
	\ifcsname fn@#1@PL\endcsname%
	\ifAC@uppercase@first%
	\expandafter\expandafter\expandafter\MakeUppercase\csname fn@#1@PL\endcsname%
	\else%
	\csname fn@#1@PL\endcsname%
	\fi%
	\else%
	\AC@acl{#1}s%
	\fi%
}%
\def\Acp#1{\AC@uppercase@firsttrue\acp{#1}\AC@uppercase@firstfalse}%
\def\AC@acp#1{%
	\ifcsname fn@#1@PL\endcsname%
	\ifAC@uppercase@first%
	\expandafter\expandafter\expandafter\MakeUppercase\csname fn@#1@PL\endcsname%
	\else%
	\csname fn@#1@PL\endcsname%
	\fi%
	\else%
	\AC@ac{#1}s%
	\fi%
}%
\def\Acfp#1{\AC@uppercase@firsttrue\acfp{#1}\AC@uppercase@firstfalse}%
\def\AC@acfp#1{%
	\ifcsname fn@#1@PL\endcsname%
	\ifAC@uppercase@first%
	\expandafter\expandafter\expandafter\MakeUppercase\csname fn@#1@PL\endcsname%
	\else%
	\csname fn@#1@PL\endcsname%
	\fi%
	\else%
	\AC@acf{#1}s%
	\fi%
}%
\def\Acsp#1{\AC@uppercase@firsttrue\acsp{#1}\AC@uppercase@firstfalse}%
\def\AC@acsp#1{%
	\ifcsname fn@#1@PL\endcsname%
	\ifAC@uppercase@first%
	\expandafter\expandafter\expandafter\MakeUppercase\csname fn@#1@PL\endcsname%
	\else%
	\csname fn@#1@PL\endcsname%
	\fi%
	\else%
	\AC@acs{#1}s%
	\fi%
}%
\edef\AC@uppercase@write{\string\ifAC@uppercase@first\string\expandafter\string\MakeUppercase\string\fi\space}%
\def\AC@acrodef#1[#2]#3{%
	\@bsphack%
	\protected@write\@auxout{}{%
		\string\newacro{#1}[#2]{\AC@uppercase@write #3}%
	}\@esphack%
}%
\def\Acl#1{\AC@uppercase@firsttrue\acl{#1}\AC@uppercase@firstfalse}
\def\Acf#1{\AC@uppercase@firsttrue\acf{#1}\AC@uppercase@firstfalse}
\def\Ac#1{\AC@uppercase@firsttrue\ac{#1}\AC@uppercase@firstfalse}
\def\Acs#1{\AC@uppercase@firsttrue\acs{#1}\AC@uppercase@firstfalse}
\makeatother

\newtheorem{theorem}{Theorem}

\acrodef{OBO}{output back-off}
\acrodef{PAPR}{peak-to-average-power ratio}
\acrodef{APAC}{aperiodic auto correlation}
\acrodef{OFDM}{orthogonal frequency division multiplexing}
\acrodef{DFT}{discrete Fourier transform}
\acrodef{DC}{direct current}
\acrodef{CS}{complementary sequence}
\acrodef{GCP}{Golay complementary pair}
\acrodef{ANF}{algebraic normal form}
\acrodef{PSK}{phase shift keying}
\acrodef{QAM}{quadrature amplitude modulation}
\acrodef{QPSK}{quadrature phase shift keying}
\acrodef{GDJ}{Golay-Davis-Jedwab}
\acrodef{PMEPR}{peak-to-mean envelope power ratios}
\acrodef{FFT}{fast Fourier transform}
\acrodef{BER}{bit-error rate}
\acrodef{SNR}{signal-to-noise ratio}
\acrodef{4G}{Fourth Generation}
\acrodef{5G}{Fifth Generation}
\acrodef{NR}{5G New Radio}
\acrodef{LTE}{Long-Term Evolution}
\acrodef{PTS}{partial transmit sequences}
\acrodef{PSD}{power spectral density}
\acrodef{LDPC}{low-density parity check}
\acrodef{CP}{cyclic prefix}
\acrodef{OOK}{on-off keying}
\acrodef{WUR}{wake-up radio}
\acrodef{WUS}{wake-up signal}
\acrodef{PAPR}{peak-to-average-power ratio}
\acrodef{WFC}{waveform coding}
\acrodef{OFDM}{orthogonal frequency division multiplexing}
\acrodef{MPC}{multi-path channel}
\acrodef{GI}{guard interval}
\acrodef{RMSE}{root-mean-squared error}
\acrodef{IDFT}{inverse discrete Fourier transform}
\acrodef{DFT}{discrete Fourier transform}
\acrodef{ISL}{integrated side lobe}
\acrodef{CAN}{cyclic algorithm-new}
\acrodef{SCAN}{shaping with CAN}
\acrodef{FFT}{fast Fourier transform}
\acrodef{BER}{bit error rate}
\acrodef{AWGN}{additive white Gaussian noise}
\acrodef{QAM}{quadrature amplitude modulation}
\acrodef{SNR}{signal-to-noise ratio}
\acrodef{PDP}{power delay profile}
\acrodef{PPM}{pulse-position modulation}
\acrodef{ACI}{adjacent-channel interference}
\acrodef{PA}{power amplifier}
\acrodef{BLE}{Bluetooth Low Energy}
\acrodef{RF}{radio frequency}
\acrodef{FSK}{frequency-shift keying}
\acrodef{OFDMA}{orthogonal frequency division multiple access}
\acrodef{MAC}{medium access control}
\acrodef{WURx}{wake-up radio receiver}
\acrodef{WUTx}{wake-up radio transmitter}
\acrodef{LPF}{low-pass filter}
\acrodef{DC}{direct current}
\acrodef{IoT}{Internet-of-things}
\acrodef{LDR}{low data rate}
\acrodef{HDR}{high data rate}
\acrodef{FDMA}{frequency division multiple access}
\acrodef{LFSR}{linear feedback shift register}
\acrodef{LTF}{long training field}
\acrodef{STF}{short training field}
\acrodef{SIG}{signal field}
\acrodef{CDF}{cumulative distribution function}
\acrodef{PAM}{pulse amplitude modulation}
\acrodef{AP}{access point}
\acrodef{SEM}{spectral emission mask}
\acrodef{WLAN}{Wireless Local Area Network}
\acrodef{FCC}{Federal Communications Commission}
\begin{document}
\title{ 
Low-PAPR Multi-channel OOK Waveform for IEEE 802.11ba Wake-up Radio
}
\author{...,~\IEEEmembership{Member,~IEEE,}}
 \author{Alphan~\c{S}ahin, Xiaofei~Wang, Hanqing~Lou, and~Rui~Yang
 \thanks{Alphan~\c{S}ahin is affiliated with the University of South Carolina, Columbia, SC and Xiaofei~Wang, Hanqing~Lou, and Rui~Yang are affiliated with InterDigital, Huntington Quadrangle, Melville, NY. E-mail: asahin@mailbox.sc.edu, \{xiaofei.wang, hanqing.lou, rui.yang\}@interdigital.com}
 \vspace{-3mm}
 }
\maketitle

\pagenumbering{gobble}
\begin{abstract}

The \ac{PAPR} of the frequency domain multiplexed \acp{WUS} specified in IEEE P802.11ba can be very large and difficult to manage since it depends on the number and allocation of the active channels, and the data rate on each channel. To address this issue, we propose a transmission scheme based on \acp{CS} for multiple \acp{WUS} multiplexed in the frequency domain. We discuss how to construct \acp{CS} compatible with the framework of IEEE P802.11ba by exploiting a recursive \ac{GCP} construction to reduce the instantaneous power fluctuations in time. We compare the proposed scheme with the other options under a non-linear \ac{PA}  distortion. Numerical results show that the proposed scheme can lower the \ac{PAPR} of the transmitted signal in \ac{FDMA} scenarios more than 3~dB and yields a superior error rate performance under severe \ac{PA} distortion.

\end{abstract}

\acresetall

\section{Introduction}

A \ac{WUR} is an extremely low-power radio attached to a primary radio. It listens the wireless environment while its primary radio is in sleep mode and wakes it up when there is a packet to receive. Due to its low power consumption, it can potentially reduce the power consumption of the device to maintain the wireless connectivity for many years, which is particularly beneficial for battery-operated and remotely-connected devices \cite{McCormick_2017}. 

\ac{WUR} is being discussed under an ongoing IEEE project, i.e., IEEE P802.11ba, which aims at designing \ac{WLAN} with \ac{WUR} devices \cite{baDraft_2019}
To enable low-complexity \ac{WURx}, an \ac{OOK}-based \ac{WUS} has been adopted in IEEE P802.11ba while providing the flexibility for the implementation of the \ac{OOK} waveform to system designers \cite{baDraft_2019}. In this study, we investigate how to design the \ac{OOK}-based \ac{WUS} in \ac{FDMA} scenarios under the framework of IEEE P802.11ba.


In the literature, the design of \ac{WUS} for single channel has been investigated extensively. For example, in \cite{zhang_2018}, \ac{OOK} signals in a \ac{WUS} are generated through \ac{OFDM} symbols with high-power and low-power constellation points, respectively. In \cite{sahin_2018seq}, an approach which enables orthogonal multiplexing of \ac{WUS} and \ac{OFDM} symbol is proposed. The sequences which lead to ON and OFF signals within the useful duration of \ac{OFDM} signal are optimized based on several parameters such as \ac{PAPR}, flatness in frequency, and low leakage on the OFF signals.
In \cite{wilhelmsson2018}, simultaneous \ac{WUS} and \ac{OFDM} signal transmission is discussed under the framework of IEEE 802.11ax. To mitigate the interference between \ac{WUS} and \ac{OFDM} signal, filtered \ac{OOK} waveform is proposed. In \cite{sundman_2018},  the concept of partial \ac{OOK}, which improves the sensitivity performance of a \ac{WUR} without increasing the complexity, is investigated.
There are also other attempts to generate a low data rate \ac{WUS} by exploiting the existing IEEE 802.11 amendments. For example, in \cite{kim_2016}, special bit streams which purposely yield to \ac{OFDM} symbols  with high \acp{PAPR} in IEEE 802.11a/g/n packets are proposed. The location of the peak sample is controlled in time by using different bit streams for \ac{WUS}. 
Mapping the low-power constellation points to the lower half of the channel bandwidth, and high-power points to the upper half to encode a bit 0 (and vice versa for bit 1), i.e., \ac{FSK},  is also another strategy mentioned in \cite{kim_2016}. In \cite{tang_2015} and \cite{Roberts_2016}, the information bits related to \ac{WUS} are encoded with the frame length by using multiple consecutive Wi-Fi or \ac{BLE} packets. 
A comprehensive survey on other modulation techniques for \acp{WUR} can be found in \cite{Piyare_2017}. To the best of our knowledge, the  design of \ac{OOK} waveform in \ac{FDMA} scenarios has not been investigated rigorously yet.

One of the main challenges to design an \ac{OOK} waveform applicable in \ac{FDMA} scenarios \cite{baDraft_2019} is the large instantaneous power fluctuation. 
To address this issue, we first propose a transmission scheme which unifies the \ac{LDR} \ac{WUS} and the \ac{HDR} \ac{WUS} generation methods. We then generate \ac{OOK} waveform based on \acp{CS} to limit the \ac{PAPR} below $3$~dB. We show how to construct \acp{CS} compatible with \ac{FDMA} in IEEE P802.11ba by exploiting a recursive \ac{GCP} construction. We compare the proposed scheme with the other options provided in \cite{baDraft_2019} under non-linear \ac{PA}  distortion. 


{\em Notation:} The field of complex numbers,  the set of integers, the set of positive integers, and the set of non-negative integers  are denoted by $\complexNumbers$,  $\integers$,  $\integersPositive$, and $\integersNonnegative$ respectively.
The symbol $*$ denotes convolution operation.
The symbols $\constanti$, $\constantMinusi$, $\constantOne$, and $\constantMinusOne$ denote $\sqrt{-1}$, $-\sqrt{-1}$, $1$, and $-1$, respectively.
The Hermitian operation and the complex conjugation are denoted by $(\cdot)^{\rm H}$ and $(\cdot)^*$ respectively.
The operator $\flipConjugate[\seqGa]$ reverses the order of the elements of the sequence $\seqGa$ and applies element-wise complex conjugation.
The operation $\upsampleOp[\upsampleVarA][\seqGa]$ introduces $\upsampleVarA-1$ zero symbols between the elements of $\seqGa$.

\section{IEEE P802.11ba Framework}
\label{sec:baFramework}
In the current draft of IEEE P802.11ba \cite{baDraft_2019}, a \ac{WUS} is based on 4~MHz \ac{OOK} waveform. While the ON and OFF signal durations are set to be $4~\mu$s for \ac{LDR} \ac{WUS}, they are $2~\mu$s for \ac{HDR} \ac{WUS}, which are half of the IEEE 802.11n/ac \ac{OFDM} symbol duration. To enable non-coherent detection, the bits for the payload are encoded with a Manchester-like encoding, called {\em \ac{WUR} encoding}. The \ac{WUR} encoding is defined as $[1~0]$ and $[0~1]$ for bit 0 and bit 1 for \ac{HDR}, and  $[1~0~1~0]$ and $[0~1~0~1]$ for bit 0 and bit 1 for \ac{LDR}, respectively. Hence, the data rates for the \acp{WUS} for \ac{HDR} and \ac{LDR} correspond to  $250$~kbit/s or $62.5$~kbit/s, respectively.

The current draft of the standard provides several ways of generating the \ac{OOK} waveform. For example, using the half duration of the 802.11n/ac OFDM symbols with $12$ non-zero subcarriers and $1$ zero center tone (i.e., masked-based approach)  or using a smaller \ac{IDFT} with $6$ non-zero subcarriers and $1$ zero center tone are two of the exemplified methods to generate the ON signals  for \ac{HDR} \ac{WUS}. The \ac{LDR} \ac{WUS} is suggested to be generated by using the same signal processing operation for \ac{HDR}, i.e., \ac{OFDM} with  $12$ non-zero subcarriers and $1$ zero center tone.

A \ac{WUS} can cause spikes on certain frequency tones if identical ON signals  are used repeatedly. However, according to \ac{FCC} Part 15.247 regulations, the maximum power in any 3 kHz band should be less than 8 dBm during any time interval of continuous transmission \cite{steveBin_2018}. Hence, the coverage range of such a \ac{WUS} would be limited. To avoid this issue, IEEE P802.11ba adopts a signal randomization method where the output of \ac{IDFT} is rotated cyclically. The amount of the rotation is based on $8$ uniformly separated values within the \ac{IDFT} duration and determined with a \ac{LFSR}.

To maintain the coexistence between different IEEE 802.11 devices at the same location, the \ac{WUS} starts with the 802.11a/g preamble (legacy fields), i.e., \ac{STF}, \ac{LTF}, and a \ac{SIG}. In addition, the legacy fields are appended with two spoofing symbols, i.e., BPSK mark,  and a SYNC field  for synchronization at the \ac{WURx}, where its duration is either $64~\mu$s and $128~\mu$s for \ac{HDR} and \ac{LDR}, respectively. The SYNC field is constructed based on \ac{HDR} \ac{OOK} waveform for both \ac{HDR} and \ac{LDR} \acp{WUS}. No WUR encoding is applied for the SYNC field. A binary sequence of length 32 is utilized for the SYNC field for \ac{HDR} \ac{WUS}. The same binary sequence is repeated and inverted for the SYNC field of \ac{LDR} \ac{WUS} \cite{jiajia_2018}.

In IEEE P802.11ba, multiplexing \acp{WUS} within 20 MHz is not allowed. However, \acp{WUS} with different data rates can be multiplexed optionally within $2\times 20$ MHz or $4\times 20$ MHz where each 20 MHz channel can have a single 4 MHz \ac{WUS}. When \acp{WUS} are multiplexed in the frequency domain, the ON signals on different channels overlap in time domain. Thus, without any precaution, the multi-channel operation can cause fluctuation in time domain and decrease the coverage range. To mitigate the fluctuations, phase rotations on different channels, as described in IEEE 802.11ac \cite{ieee_2016}, are adopted. However, this may not be  an effective method as the number of active channels changes based on the encoded bits on different channels. In addition, the number of possible signals for each $4~\mu$s grows exponentially with the number of active channels in case of \ac{FDMA} due to the random cyclic shifts in time. 


\def\encodedBit[#1]{b_{#1}}
\def\indexChannel{i}
\def\ofdmSymbol{{\bf x}}
\def\outputSequence[#1]{{\bf c}_{#1}}
\def\inputSequence[#1]{{\bf s}_{#1}}
\def\sequenceLength{L}
\def\cpSize{N_{\rm cp}}
\def\idftSize{N}
\def\CPadder{{\bf A}}
\def\DFTmatrix{{{\bf F}}_{\idftSize}}
\def\mappingMatrix{{\bf M}}
\section{Preliminaries and Further Notation}
\label{sec:pre}
We define the polynomial representation of the sequence $\seqGa$ of length $\lengthGaGb$ as 
\begin{align}
\polySeq[\seqGa][\polyVariable] \triangleq \eleGa[\lengthGaGb-1]\polyVariable^{\lengthGaGb-1} + \eleGa[\lengthGaGb-2]\polyVariable^{\lengthGaGb-2}+ \dots + \eleGa[0]~,
\end{align}
where $\polyVariable\in \complexNumbers$ is a complex number. 
The polynomial $\polySeq[\seqGa][\polyVariable^\upsampleVarA]$,  $\polySeq[\seqGa][\polyVariable^\upsampleVarA]\polySeq[\seqGb][\polyVariable^\upsampleVarB]$, and $\polySeq[\seqGa][\polyVariable]\polyVariable^\separationVar$ represent the up-sampled sequence $\seqGa$ with the factor of $\upsampleVarA\in\integersPositive$,
the convolution of the up-sampled sequence $\seqGa$ by $\upsampleVarA$ and the up-sampled sequence $\seqGb$ by $\upsampleVarB\in\integersPositive$, and   the sequence $\seqGa$ padded with $\separationVar\in\integersNonnegative$ zero symbols, respectively. The polynomial $\polySeq[\seqGa][\polyVariable]$ corresponds to an \ac{OFDM} symbol in continuous time when $\polyVariable$ is restricted to be on the unit circle in the complex plane, i.e., $\polyVariable\in\{\constante^{\constantj\frac{2\pi\timeVar}{\symbolDuration}}| 0\le\timeVar <\symbolDuration
\}$, where the subcarriers are modulated with the elements of the sequence $\seqGa$, and $\symbolDuration$ denotes the \ac{OFDM} symbol duration. 

The sequence pair  $(\seqGa,\seqGb)$ is called a \ac{GCP} if
\begin{align}
\polySeq[\seqGa][\polyVariable]\polySeq[\seqGa^*][\polyVariable^{-1}]+\polySeq[\seqGb][\polyVariable]\polySeq[\seqGb^*][\polyVariable^{-1}]=\apac[\seqGa][0]+\apac[\seqGb][0]~,
\label{eq:zDomainGCP}
\end{align}
where  $\apac[\seqGa][\lagForCorrelation]$ and $\apac[\seqGb][\lagForCorrelation]$  is the \ac{APAC} coefficient of the sequence $\seqGa$ and $\seqGb$, respectively. Each sequence in a {GCP} is a \ac{CS} \cite{Golay_1961}. The \acp{CS} are first proposed in \cite{Golay_1961} in 1961. The \ac{CS} construction methods have been extensively studied in the literature (see \cite{parker_2003} and the references therein). 
One of the most appealing features of a \ac{CS} is the low instantaneous peak power when the sequence is transmitted with an \ac{OFDM} symbol. 
If $\seqGa$ is also a unimodular \ac{CS}, i.e., $|\eleGa[i]|=|\eleGa[j]|$ for $i\neq j$, one can show that the \ac{PAPR} of the corresponding \ac{OFDM} symbols is less than or equal to 2.

In \cite{sahinRui2019ICC}, a generalized version of Golay's concatenation and interleaving methods is restated by the following theorem:
\begin{theorem}
	\label{th:golayIterative}
	Let $(\seqGa,\seqGb)$ and $(\seqGc,\seqGd)$ be \acp{GCP} of length $\lengthGaGb$ and $\lengthGcGd$, respectively, and $\angleGolay[1],\angleGolay[2]\in \{u:u\in\complexNumbers, |u|=1\}$  and $\upsampleVarA, \upsampleVarB, \separationVar\in\integers$. Then, the sequences $\seqGf$ and $\seqGg$ with their polynomial representations given by
	\begin{align}
	\polySeq[\seqGf][\polyVariable] =& \angleGolay[1]\polySeq[{\seqGa}][\polyVariable^\upsampleVarA]\polySeq[{\seqGc}][\polyVariable^\upsampleVarB]   
	+ \angleGolay[2]\polySeq[{\seqGb}][\polyVariable^\upsampleVarA]\polySeq[{\seqGd}][\polyVariable^\upsampleVarB]  \polyVariable^{\separationVar} ~, 
	\label{eq:gcpGf}
	\\
	\polySeq[\seqGg][\polyVariable] =& \angleGolay[1]\polySeq[{\seqGa}][\polyVariable^\upsampleVarA]\polySeq[{\flipConjugate[\seqGd]}][\polyVariable^\upsampleVarB] 
	- \angleGolay[2]\polySeq[{\seqGb}][\polyVariable^\upsampleVarA]\polySeq[{\flipConjugate[\seqGc]}][\polyVariable^\upsampleVarB]  \polyVariable^{\separationVar} ~, 
	\label{eq:gcpGg}
	\end{align}
	construct a \ac{GCP}.
\end{theorem}
The proof for Theorem~\ref{th:golayIterative} is given in \cite{sahinRui2019ICC} by using the definition given in \eqref{eq:zDomainGCP}. 

\section{Golay-Based Multi-Channel OOK Waveform}
\label{sec:golayBasedOOK}

In this section, we discuss a simple transmitter structure for generating ON signals on different channels for both \ac{HDR} and \ac{LDR} \acp{WUS}. By using this structure and exploiting \acp{GCP}, we show how to control the \ac{PAPR} of the signal regardless of the data rate of \ac{WUS} on different channels and signal randomization discussed in Section~\ref{sec:baFramework}.

\subsection{Aligning the OFDM symbol boundaries for LDR and HDR }
\begin{figure}[]
	\centering
	{\includegraphics[width =3.2in]{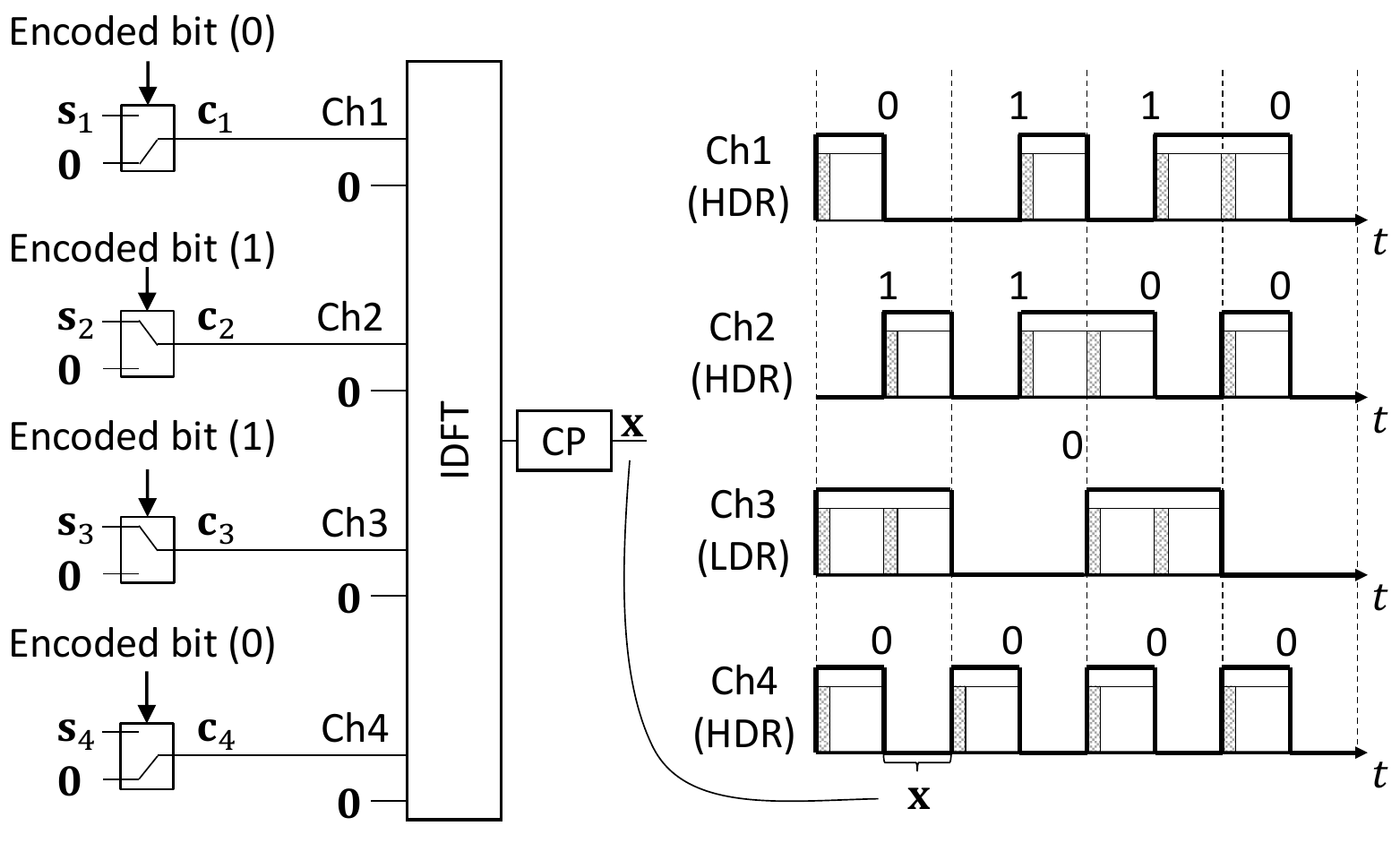}
	}
	\caption{Supporting  \ac{LDR} and \ac{HDR} \acp{WUS}  with single OFDM transmitter.}
	\label{fig:alignment}
\end{figure}

To reduce the number of possible signals for each $4~\mu$s in case of \ac{HDR} and \ac{LDR} \acp{WUS} multiplexing, we first align the \ac{OFDM} symbol periods for \ac{LDR} and \ac{HDR} by constructing each ON duration of the \ac{LDR} \ac{WUS} with two \ac{HDR} ON signals transmitted back-to-back. For example, consider a multi-channel operation where the first two channels and the fourth channel carry \ac{HDR} \ac{WUS} and the third channel carries \ac{LDR} \ac{WUS} as in \figurename~\ref{fig:alignment}. 
Each ON signal for \ac{LDR} \ac{WUS} on the third channel consists of two \ac{HDR} ON signal, in which the second OFDM symbol can be a repetition of the first OFDM symbol or another OFDM symbol generated through a different sequence. As shown in \figurename~\ref{fig:alignment}, this approach aligns the symbol boundaries of the \ac{OFDM} symbols in time for different channels. The alignment is beneficial to adjust the sequences on each channel to control the \ac{PAPR} level of the combined signal in case of \ac{FDMA} transmission of mixed \ac{LDR} and \ac{HDR} \acp{WUS}. In addition, the random cyclic shifts at the \ac{IDFT} output do not change the \ac{PAPR} properties of the generated signal. Note that this approach can also be expressed as a different \ac{WUR} encoding for \ac{LDR} case.  For example, bit 0 and bit 1  may be encoded as [1 1 0 0 1 1 0 0] and [0 0 1 1 0 0 1 1] for \ac{LDR} before the \ac{OOK} waveform generation, respectively. Hence, the ON and OFF durations for \ac{LDR} are set effectively to the ones used for \ac{HDR}.

Let $\encodedBit[\indexChannel]$ be the encoded bit and $\outputSequence[\indexChannel]\in\complexNumbers^{\sequenceLength\times 1}$ be the sequence utilized  in frequency for the $\indexChannel$th channel. If $\encodedBit[\indexChannel]=1$, $\outputSequence[\indexChannel]=\inputSequence[\indexChannel]\in\complexNumbers^{\sequenceLength\times 1}$, otherwise it is set to a zero vector. Considering four channels, the \ac{OFDM} symbol in the baseband $\ofdmSymbol\in \complexNumbers^{\idftSize\times 1}$ can be expressed as
$\ofdmSymbol =\CPadder \DFTmatrix^{\rm H} \mappingMatrix[\outputSequence[1]^{\rm H} ~ \outputSequence[2]^{\rm H}~ \outputSequence[3]^{\rm H} ~ \outputSequence[4]^{\rm H}]^{\rm H}
$,
where $\CPadder\in \complexNumbers^{\idftSize+\cpSize\times\idftSize}$ is a matrix that appends the \ac{CP} to the beginning of the symbol, $\DFTmatrix\in \complexNumbers^{\idftSize\times\idftSize}$ is the \ac{DFT} matrix, and $\mappingMatrix\in\complexNumbers^{\idftSize\times4\sequenceLength}$ is the mapping matrix that maps $\outputSequence[\indexChannel]$ to the $\indexChannel$th channel for $\indexChannel=1,2,3,4$. The transmitter diagram is illustrated in \figurename~\ref{fig:alignment}. In one implementation, $\idftSize$ and $\cpSize$   can be set to  $128$ and $16$ for 80 MHz sample rate. Hence, $\ofdmSymbol$ will be transmitted within $2~\mu$s, which is equivalent to ON duration for \ac{HDR} \ac{WUS}. If  $\outputSequence[\indexChannel]=\inputSequence[\indexChannel]$, the $\indexChannel$th channel is activated for $2~\mu$s. If it is activated twice, the ON duration for \ac{LDR} \ac{WUS} is generated on $\indexChannel$th channel. 
In the following subsection, we discuss the design of $\inputSequence[\indexChannel]$ for $\indexChannel=1,2,3,4$.

\subsection{Golay-based Multi-channel OOK Waveform}
The parameters $\upsampleVarA$ and $\upsampleVarB$ upsample the sequences in Theorem~\ref{th:golayIterative} while the parameter $\separationVar$ separates the corresponding sequences by padding zeros. Since the parameter $\separationVar$ can be an arbitrary value, Theorem~\ref{th:golayIterative} gives flexibility to control the \ac{PAPR} of a signal for non-contiguous resource allocation. In this study, we exploit this property to mitigate the \ac{PAPR} of a signal which consists of multiple ON signals in different channels while maintaining zero center tone (e.g., DC tone) on each channel.

Let $\seqGa$ and $\seqGb$ be a \ac{GCP} of length $\lengthGaGb$, $\seqGc=\seqGd=1$, $\upsampleVarA=\upsampleVarB=1$, and $\separationVar=\lengthGaGb+1$. Based on \eqref{eq:gcpGf} and \eqref{eq:gcpGg}, the corresponding sequences in the \acp{GCP} can be obtained as
\begin{align}
\seqGf_1 =&
\begin{pmatrix}
\angleGolay[1]\times\seqGa,0,\angleGolay[2]\times\seqGb
\end{pmatrix}
\label{eq:gcpP1f}
\\
\seqGg_1 =& 
\begin{pmatrix}
\angleGolay[1]\times\seqGa,0,-\angleGolay[2]\times\seqGb
\end{pmatrix}
\label{eq:gcpP1g}
\end{align}
Similarly, by alternating the variables in Theorem~\ref{th:golayIterative},
$
\seqGf =
(
\angleGolay[1]\times\seqGa,0,\angleGolay[2]\times\seqGb
)
$
and
$
\seqGg =
(
\angleGolay[1]\times\tilde{\seqGa},0,-\angleGolay[2]\times\tilde{\seqGb}.
)
$
also construct a \ac{GCP}. Hence, by choosing $\lengthGaGb=3$ and $\angleGolay[1]=\angleGolay[2]=1$, a set of \acp{CS} with zero center tone and compatible with IEEE P802.11ba \ac{OOK} waveform can be generated. For example, let $\seqGa = (+,\constanti,+)$ and $\seqGb=(+,+,-)$ based on the \ac{GCP} lists provided in \cite{holzmann_1991}. By utilizing \eqref{eq:gcpP1f} and \eqref{eq:gcpP1g}, $\seqGf_1$ and $\seqGg_1$ are obtained as $\seqGf_1=(+,\constanti,+,0,+,+,-)$ and $\seqGg_1=(+,\constanti,+,0,-,-,+)$ and they can be used for generating $2~\mu$s ON signal with the  transmitter structure in \figurename~\ref{fig:alignment}. 
To utilize $\seqGf_1$ and $\seqGg_1$  for multi-channel operations, the following rules may be considered:

\subsubsection{Case I (The number of active channels is 1)}
When only one of the channels is active, either $\seqGf_1= (\seqGa,0,\seqGb)$ and $\seqGg_1= (\seqGa,0,-\seqGb)$  can be mapped to the inputs of \ac{IDFT} to generate ON signal for the corresponding channel. Since $\seqGf_1$ and $\seqGg_1$ are the unimodular \acp{CS}, they lead to the \ac{OFDM} signals where their \acp{PAPR} are less than 3 dB. 

\subsubsection{Case II (The number of active channels is 2)}
The sequences $\seqGf_1$ and $\seqGg_1$ construct \ac{GCP}. Therefore, $\seqGf_1$ and $\seqGg_1$ can be re-used in  \eqref{eq:gcpGf} and \eqref{eq:gcpGg} to form larger \acp{CS}. For $\angleGolay[1]=\angleGolay[2]=1$, the new pair can be obtained as
\begin{align}
\seqGf_2 =&
\begin{pmatrix}
\seqGf_1,{\bf 0}_{\separationVar},\seqGg_1
\end{pmatrix}
=
\begin{pmatrix}
\seqGa,0,\seqGb,{\bf 0}_{\separationVar},
\seqGa,0,-\seqGb
\end{pmatrix}~,
\label{eq:gcpP1ff}
\\
\seqGg_2 =& 
\begin{pmatrix}
\seqGf_1,{\bf 0}_{\separationVar},-\seqGg_1
\end{pmatrix}
=
\begin{pmatrix}
\seqGa,0,-\seqGb,{\bf 0}_{\separationVar},
-\seqGa,0,\seqGb
\end{pmatrix}~.
\label{eq:gcpP1gg}
\end{align}
by using the same rationale in \eqref{eq:gcpP1f} and \eqref{eq:gcpP1g}, where $\separationVar$ can be chosen arbitrarily. In other words,  the sequences $\seqGf_1$ and $\seqGg_1$ can be mapped to different channels regardless of the separation between the channels in frequency without affecting the \ac{PAPR}. For example, the sequences $\seqGf_1$ and $\seqGg_1$  can be utilized to generate ON signals for any two of the four channels. Note that $\seqGf_1$ and $\seqGg_1$ can also be manipulated based on Theorem~\ref{th:golayIterative} before mapping to the subcarriers. For example, one channel can utilize the sequence $\constantj\times \seqGf_1$  while the other channel may utilize the sequence $\constanti\times\seqGg_1$, i.e.,  $\angleGolay[1]=\constantj$ and $\angleGolay[2]=\constanti$.

\subsubsection{Case III (The number of active channels is 3)}
For this case, either three contiguous channels or the first, second (or third), and the fourth channels are active at the same time. To address the first scenario, let $\upsampleVarB=1$, $\seqGc=\seqGa$, and $\seqGd=\seqGb$. The generated sequences based on Theorem~\ref{th:golayIterative} can  be expressed as  $\seqGf_3 = (\upsampleOp[\upsampleVarA][\seqGa]*\seqGa, {\bf 0}_\separationVar) + ({\bf 0}_\separationVar, \upsampleOp[\upsampleVarA][\seqGb]*\seqGb) $ and $\seqGg_3 = (\upsampleOp[\upsampleVarA][\seqGa]*\tilde{\seqGb}, {\bf 0}_m) - ({\bf 0}_m, \upsampleOp[\upsampleVarA][\seqGb]*\tilde{\seqGa})$. If $\separationVar=\lengthGaGb+1$ and $\upsampleVarA\ge2\lengthGaGb+1$, the sequences which cover three contiguous channels can be formed by using a \ac{GCP} of length 3. For example,  if $\seqGa = (+,\constanti,+)$ and $\seqGb=(+,+,-)$, for $\angleGolay[1]=\angleGolay[2]=1$, $\seqGf_3$ and $\seqGg_3$ can be obtained as
\begin{align}
\seqGf_3 =&
\begin{pmatrix}
\seqGa,0,\seqGb,{\bf 0}_{\upsampleVarA-2\lengthGaGb-1},
i\seqGa,0,\seqGb,{\bf 0}_{\upsampleVarA-2\lengthGaGb-1},
\seqGa,0,-\seqGb
\end{pmatrix}~,
\label{eq:gcpP1fff}
\\
\seqGg_3 =& 
\begin{pmatrix}
\seqGa,0,-\seqGb,{\bf 0}_{\upsampleVarA-2\lengthGaGb-1},
i\seqGa,0,-\seqGb,{\bf 0}_{\upsampleVarA-2\lengthGaGb-1},
\seqGa,0,\seqGb
\end{pmatrix}.
\label{eq:gcpP1ggg}
\end{align}
By using \eqref{eq:gcpP1fff}, the ON duration for the first, the second, and the third channel can be generated by using $(\seqGa,0,\seqGb)$, $(\constanti\seqGa,0,\seqGb)$, and $(\seqGa,0,-\seqGb)$ (or $(\seqGa,0,-\seqGb)$, $(\constanti\seqGa,0,-\seqGb)$, and $(\seqGa,0,\seqGb)$ based on \eqref{eq:gcpP1ggg}), respectively. Unfortunately, to the best of our knowledge, Theorem~\ref{th:golayIterative} does not yield \ac{CS} with \ac{QPSK} alphabet when the active channels are not contiguous. In this study, we use $(\seqGa,0,\seqGb)$ and $(\seqGa,0,-\seqGb)$  for the first and second channel, respectively, and choose $\angleGolay[1]$ and $\angleGolay[2]$ from \ac{QPSK} alphabet for the sequence $(\angleGolay[1]\times\seqGa,0,-\angleGolay[2]\times\seqGb)$  for the third (or the second) channel to minimize \ac{PAPR} while maintaining the dynamic range for each individual channel low and decreasing the transmitter complexity.

\subsubsection{Case IV (The number of active channels is 4)}
This case can be addressed by utilizing Theorem~\ref{th:golayIterative} and exploiting the \ac{GCP} of $\seqGf_2$ and $\seqGg_2$. For $\angleGolay[1]=\angleGolay[2]=1$, the sequences can be obtained as
\begin{align}
\seqGf_3 =&
\begin{pmatrix}
\seqGf_2,{\bf 0}_{\separationVar},\seqGg_2
\end{pmatrix}
=
\begin{pmatrix}
\seqGf_1,{\bf 0}_{\separationVar},\seqGg_1,{\bf 0}_{\separationVar},
\seqGf_1,{\bf 0}_{\separationVar},-\seqGg_1
\end{pmatrix}~,
\label{eq:gcpP1ffff}
\\
\seqGg_3 =& 
\begin{pmatrix}
\seqGf_2,{\bf 0}_{\separationVar},-\seqGg_2
\end{pmatrix}
=
\begin{pmatrix}
\seqGf_1,{\bf 0}_{\separationVar},\seqGg_1,{\bf 0}_{\separationVar},
-\seqGf_1,{\bf 0}_{\separationVar},\seqGg_1
\end{pmatrix}~.
\label{eq:gcpP1gggg}
\end{align}
Thus, the first, the second, the third, and the forth channels can utilize $\seqGf_1$, $\seqGg_1$, $\seqGf_1$ and $-\seqGg_1$ (or $\seqGf_1$, $\seqGg_1$, $-\seqGf_1$ and $\seqGg_1$), respectively,  as indicated in \eqref{eq:gcpP1fff}, to maintain the \ac{PAPR}  less than $3$~dB even though the four channels are active simultaneously.

\begin{table}[]
	\caption{Sequences for Multi-channel OOK Waveform}
	\label{tab:seq}
	\centering
	\begin{tabular}{|c|c|c|c|c|}
		\hline
		$\encodedBit[1],\encodedBit[2],\encodedBit[3],\encodedBit[4]$ & $\outputSequence[1]$ & $\outputSequence[2]$ & $\outputSequence[3]$ & $\outputSequence[4]$ \\ \hline
		$0000$                                                        &               ${\bf 0}_{\sequenceLength}$       &          ${\bf 0}_{\sequenceLength}$           &      ${\bf 0}_{\sequenceLength}$                &         ${\bf 0}_{\sequenceLength}$             \\ \hline
		$1000$                                                        &         $(\seqGa,0,\seqGb)$            &         ${\bf 0}_{\sequenceLength}$             &          ${\bf 0}_{\sequenceLength}$            &          ${\bf 0}_{\sequenceLength}$            \\ \hline
		$0100$                                                        &            ${\bf 0}_{\sequenceLength}$           &       $(\seqGa,0,\seqGb)$               &         ${\bf 0}_{\sequenceLength}$              &          ${\bf 0}_{\sequenceLength}$             \\ \hline
		$1100$                                                        &       $(\seqGa,0,\seqGb)$                &          $(\seqGa,0,-\seqGb)$             &            ${\bf 0}_{\sequenceLength}$           &            ${\bf 0}_{\sequenceLength}$           \\ \hline
		$0010$                                                        &           ${\bf 0}_{\sequenceLength}$             &              ${\bf 0}_{\sequenceLength}$          &      $(\seqGa,0,-\seqGb)$                 &          ${\bf 0}_{\sequenceLength}$              \\ \hline
		$1010$                                                        &          $(\seqGa,0,\seqGb)$            &       ${\bf 0}_{\sequenceLength}$              &           $(\seqGa,0,-\seqGb)$           &     ${\bf 0}_{\sequenceLength}$                  \\ \hline
		$0110$                                                        &      ${\bf 0}_{\sequenceLength}$               &        $(\seqGa,0,\seqGb)$              &    $(\seqGa,0,-\seqGb)$                  &   ${\bf 0}_{\sequenceLength}$                    \\ \hline
		$1110$                                                        &       $(\seqGa,0,\seqGb)$               &        $(i\seqGa,0,\seqGb)$              &    $(\seqGa,0,-\seqGb)$                  &          ${\bf 0}_{\sequenceLength}$             \\ \hline
		$0001$                                                        &     ${\bf 0}_{\sequenceLength}$                &   ${\bf 0}_{\sequenceLength}$                  &       ${\bf 0}_{\sequenceLength}$               &   $(\seqGa,0,\seqGb)$                     \\ \hline
		$1001$                                                        &        $(\seqGa,0,\seqGb)$              &      ${\bf 0}_{\sequenceLength}$               &   ${\bf 0}_{\sequenceLength}$                   &    $(\seqGa,0,-\seqGb)$                    \\ \hline
		$0101$                                                        &      ${\bf 0}_{\sequenceLength}$               &     $(\seqGa,0,\seqGb)$                 &    ${\bf 0}_{\sequenceLength}$                  &  $(\seqGa,0,-\seqGb)$                      \\ \hline
		$1101^*$                                                        &         $(\seqGa,0,\seqGb)$             &      $(\constanti\seqGa,0,\seqGb)$                &              ${\bf 0}_{\sequenceLength}$        &   $(\seqGa,0,-\seqGb)$                     \\ \hline
		$0011$                                                        &        ${\bf 0}_{\sequenceLength}$             &    ${\bf 0}_{\sequenceLength}$                 &      $(\seqGa,0,\seqGb)$                &     $(\seqGa,0,-\seqGb)$                   \\ \hline
		$1011^*$                                                        &           $(\seqGa,0,\seqGb)$           &       ${\bf 0}_{\sequenceLength}$              &    $(\constantj\seqGa,0,-\seqGb)$                  &        $(\seqGa,0,-\seqGb)$                \\ \hline
		$0111$                                                        &            ${\bf 0}_{\sequenceLength}$         &          $(\seqGa,0,\seqGb)$           &            $(i\seqGa,0,\seqGb)$         &         $(\seqGa,0,-\seqGb)$            \\ \hline
		$1111$                                                        &            $(\seqGa,0,\seqGb)$         &          $(\seqGa,0,-\seqGb)$           &            $(-\seqGa,0,-\seqGb)$         &         $(-\seqGa,0,\seqGb)$            \\ \hline
	\end{tabular}
	\vspace{-2mm}
\end{table}

In \tablename~\ref{tab:seq}, the sequences for given active channels in one \ac{OFDM} symbol duration are listed. Other than the cases indicated by $^*$ in \tablename~\ref{tab:seq}, i.e.,  $(\encodedBit[1],\encodedBit[2],\encodedBit[3],\encodedBit[4]) = (1,1,0,1)$ and $(\encodedBit[1],\encodedBit[2],\encodedBit[3],\encodedBit[4]) = (1,0,1,1)$, the sequences $\outputSequence[\indexChannel]$ for $\indexChannel=1,2,3,4$ form a \ac{CS} when they mapped to the channels and the \ac{PAPR} of the corresponding signal is less than or equal to $3$~dB. In addition, the dynamic range for each channel is also limited since $\outputSequence[\indexChannel]$ itself is a \ac{CS} for $\indexChannel=1,2,3,4$. Another observation from \tablename~\ref{tab:seq} is that 
the sequence in one channel changes based on the transmitted bits on the other channels. However, since \ac{WURx} is expected to be a simple \ac{OOK} receiver, e.g., envelope detector, this is not an issue for \ac{WURx}.

\section{Numerical Results}
\label{sec:numerical}
We set the sampling rate to $320$~MHz for the simulation, which is sufficient to reveal time-domain characteristics. For the \ac{PA} model, we consider Rapp model with a smoothness factor of 3. For \ac{WURx}, we consider an envelope detector where the received samples are accumulated  as the sum of absolutes of I- and Q-branches. We utilize a $5$th order Butterworth  low-pass filter  as the receive filter whose bandwidth is $5$~MHz. We assume ideal synchronization. 
For the proposed scheme (Golay-based), we consider the \acp{CS} given in \tablename~\ref{tab:seq}, where $\seqGa = (+,\constanti,+)$ and $\seqGb=(+,+,-)$ \cite{holzmann_1991}.
We compare the proposed scheme with the three examples provided and denoted as Example 1-3 in \cite{baDraft_2019}.
For Example 1-3, the length 7 sequences and the length 13 sequences are mapped to the corresponding inputs of \ac{IDFT}.
The $2~\mu$s and $4~\mu$s ON signals for \ac{HDR} and \ac{LDR} are generated after $0.4~\mu$s and $0.8~\mu$s \ac{CP} are appended to $1.6~\mu$s and $3.2~\mu$s \ac{IDFT} outputs, respectively. We also apply phase rotations for the channels as described in IEEE 802.11ac \cite{ieee_2016} for Example 2-4. The phase rotations are not applied for the proposed method. To avoid spikes in frequency due to the \ac{OOK} waveform, we consider time domain cyclic shifts for \ac{HDR} and \ac{LDR} ON signals, where the amount of the cyclic shifts are chosen randomly from 8 uniformly separated values within the $1.6~\mu$s and $3.2~\mu$s \ac{IDFT} durations, respectively. For the proposed method, we cyclically shift the output of \ac{IDFT}.
\begin{figure}[]
	\centering
	{\includegraphics[width =3.4in]{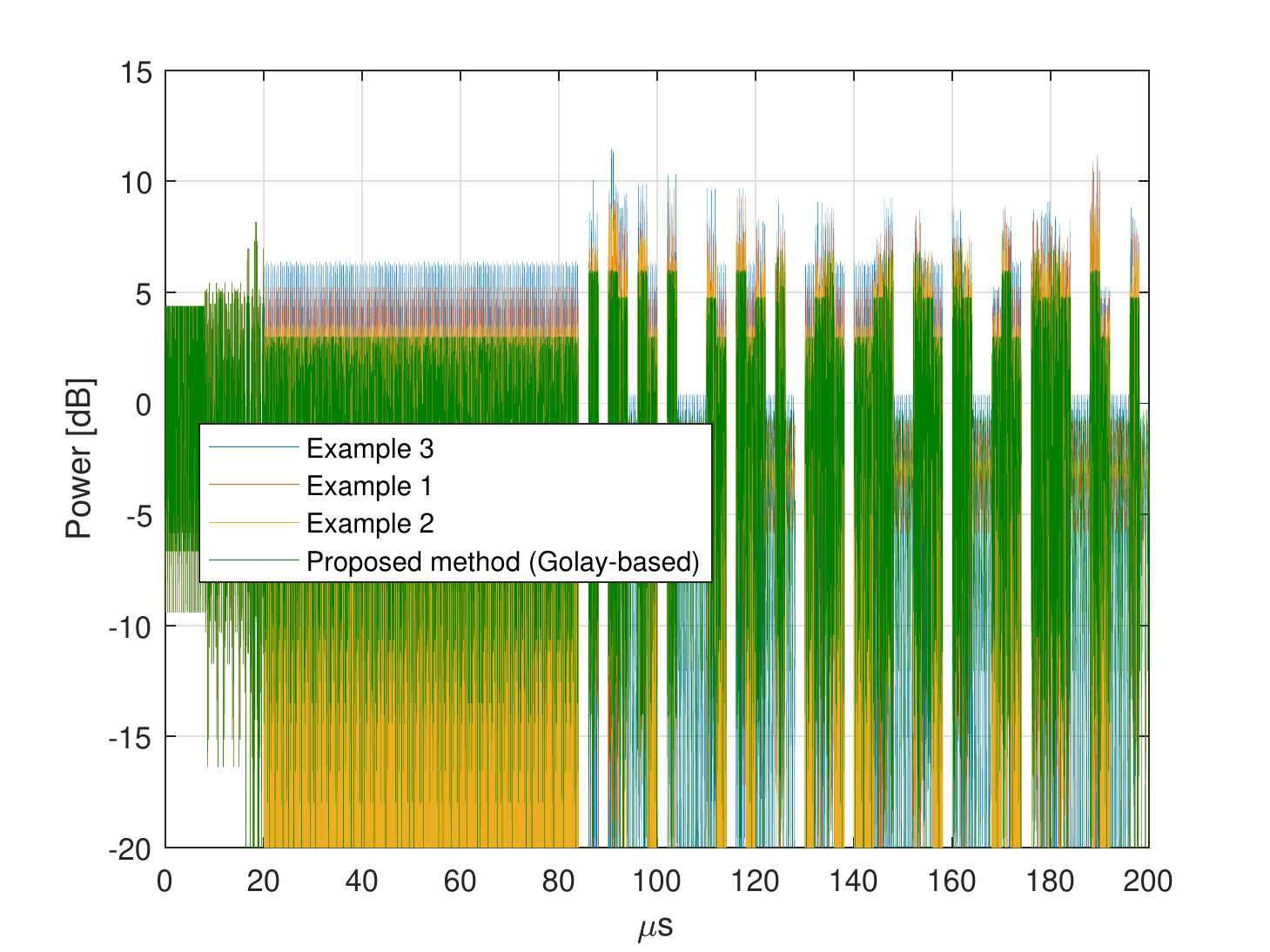}
	}
	\caption{Temporal characteristics of IEEE P802.11ba signal when two \ac{LDR} and two \ac{HDR} \acp{WUS} are multiplexed in frequency.}
	\label{fig:temp}
	\vspace{-2mm}
\end{figure}

\subsection{Temporal Characteristics}
In \figurename~\ref{fig:temp}, we provide the snapshots for IEEE P802.11ba packets generated through different options.  In this analysis, while the first two channels carry \ac{HDR} \acp{WUS}, the last two channels are loaded with two \ac{LDR} \acp{WUS}. The average power is set to 1. The first 20 $\mu$s consist of the legacy fields. The following $64~\mu$s is the combination of the SYNC fields for \ac{LDR} and \ac{HDR} \acp{WUS}. The subsequent $64~\mu$s includes the combination of the \ac{OOK} signals at the SYNC fields for the \ac{LDR} and \ac{HDR} \acp{WUS}. \figurename~\ref{fig:temp} shows that the instantaneous power for Example 2-4 can be $11-11.5$~dB and $4-5$~dB higher than the average power and the instantaneous power for  the legacy header, respectively. We observe that   Example 3  causes the highest fluctuations, whereas the proposed method mitigates them substantially.  \figurename~\ref{fig:temp} also shows that the instantaneous power level with the proposed method is similar to that of the legacy \ac{SIG} and random cyclic shifts do not alter the variation of the signals.

\subsection{PAPR Results and Spectral Regrowth}
\begin{figure}[t]
	\centering
	{\includegraphics[width =3.4in]{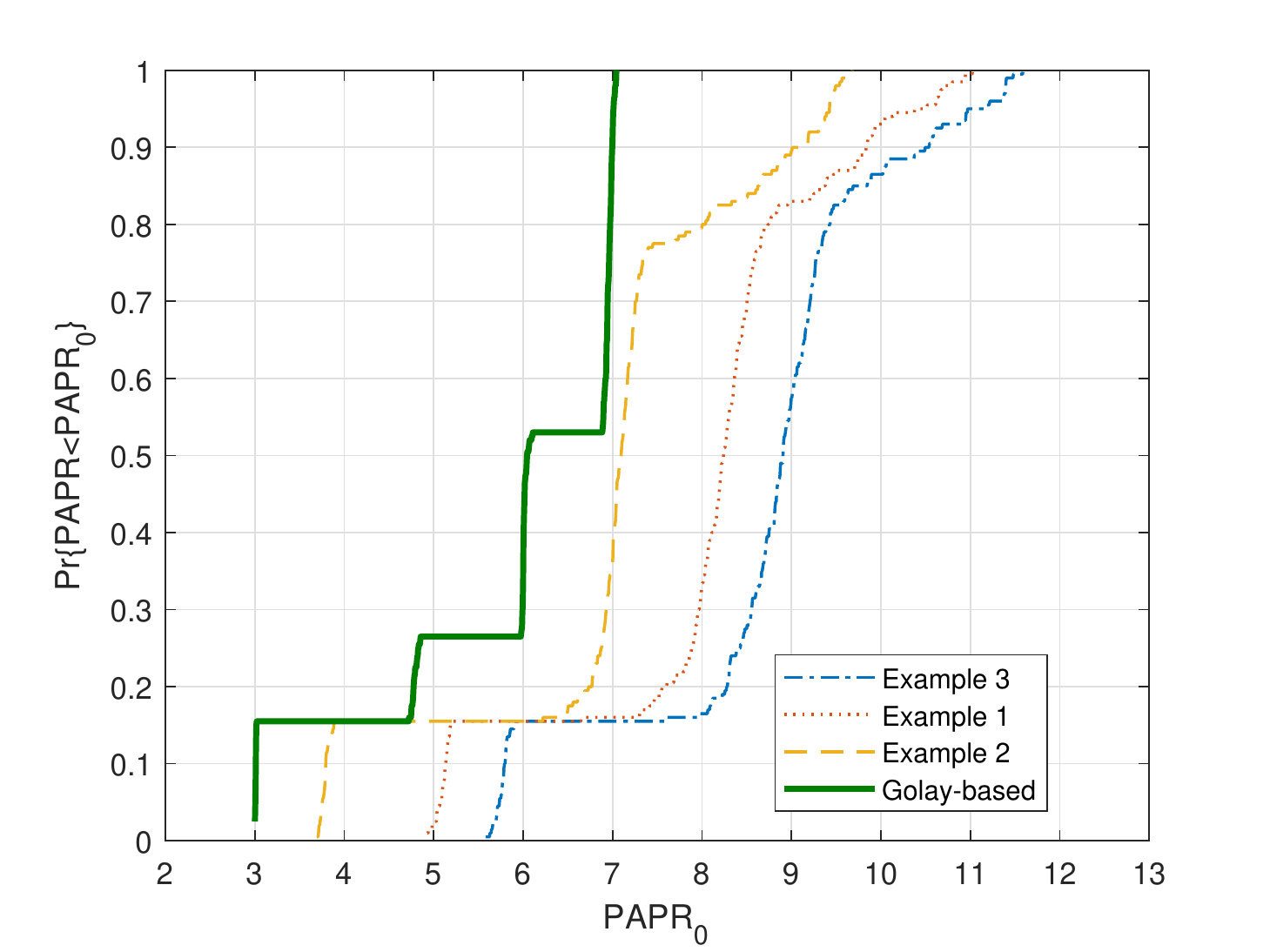}
	}
	\caption{PAPR distribution.}
	\label{fig:papr}
	\vspace{-2mm}
\end{figure}

In \figurename~\ref{fig:papr}, we compare the \ac{PAPR} distributions for different options. We consider the same setup as given for  \figurename~\ref{fig:temp}, i.e., two  \ac{HDR} \acp{WUS} and two \ac{LDR} \acp{WUS}. We measure the \ac{PAPR} for each $4~\mu$s. The results are aligned with the temporal characteristics given in \figurename~\ref{fig:temp}. While the \acp{PAPR} can reach to $9.5$, $11.5$ and $11$ dB for Example 2, Example 3, and Example 1, respectively, the \ac{PAPR} is limited to $7$ dB in the worst case for the proposed method. In other words, the proposed method provides $3.5$, $4.5$, and $4$~dB \ac{PAPR} gain as compared to Example 2, Example 3, and Example 1, respectively, for this scenario.  In \tablename~\ref{tab:allPAPR}, we also provide \ac{PAPR} at \%50, \%80, and \%99 percentiles at \ac{CDF} curves for different configurations. The results show gains in \ac{PAPR} for all scenarios.
\begin{table*}[]
\centering
\caption{PAPR performance at different percentiles for different scenarios (H: HDR, L: LDR, X: None)}
\begin{tabular}{l|c|c|c|c|c|c|c|c|c|c|c|c|c|c|c|}
\cline{2-16}
\multicolumn{1}{c|}{}             & \multicolumn{3}{c|}{\textbf{Case I {[}dB{]}}} & \multicolumn{3}{c|}{\textbf{Case II {[}dB{]}}} & \multicolumn{3}{c|}{\textbf{Case III {[}dB{]}}} & \multicolumn{3}{c|}{\textbf{Case IV {[}dB{]}}} & \multicolumn{3}{c|}{\textbf{Case V {[}dB{]}}} \\ \cline{2-16} 
\multicolumn{1}{r|}{Rates:}       & \multicolumn{3}{c|}{X X H H}                  & \multicolumn{3}{c|}{H H H H}                   & \multicolumn{3}{c|}{L L H H}                    & \multicolumn{3}{c|}{L L L H}                   & \multicolumn{3}{c|}{H H L H}                  \\ \cline{2-16} 
\multicolumn{1}{r|}{Percentile:}  & 0.5           & 0.8           & 0.99          & 0.5           & 0.8            & 0.99          & 0.5            & 0.8           & 0.99           & 0.5           & 0.8           & 0.99           & 0.5           & 0.8           & 0.99          \\ \hline
\multicolumn{1}{|l|}{Golay-based} & 6.02          & 6.10          & 6.19          & 6.01          & 6.97           & 7.05          & 6.04           & 6.97          & 7.04           & 6.92          & 6.98          & 7.03           & 6.06          & 6.96          & 7.03          \\ \hline
\multicolumn{1}{|l|}{Example 2}   & 6.58          & 6.59          & 6.61          & 7.09          & 7.10           & 7.13          & 7.09           & 8.04          & 9.61           & 7.22          & 9.25          & 9.87           & 7.09          & 9.14          & 9.60          \\ \hline
\multicolumn{1}{|l|}{Example 3}  & 9.31          & 9.42          & 9.55          & 9.86          & 12.34          & 12.53         & 8.90           & 9.43          & 11.48          & 8.71          & 9.41          & 11.30          & 9.40          & 10.98         & 11.89         \\ \hline
\multicolumn{1}{|l|}{Example 1}   & 8.21          & 8.32          & 8.49          & 8.73          & 8.83           & 8.94          & 8.24           & 8.74          & 10.94          & 8.11          & 9.37          & 11.00          & 8.63          & 10.09         & 11.17         \\ \hline
\end{tabular}
\label{tab:allPAPR}
\vspace{-7mm}
\end{table*}

\begin{figure}[t]
	\centering
	{\includegraphics[width =3.4in]{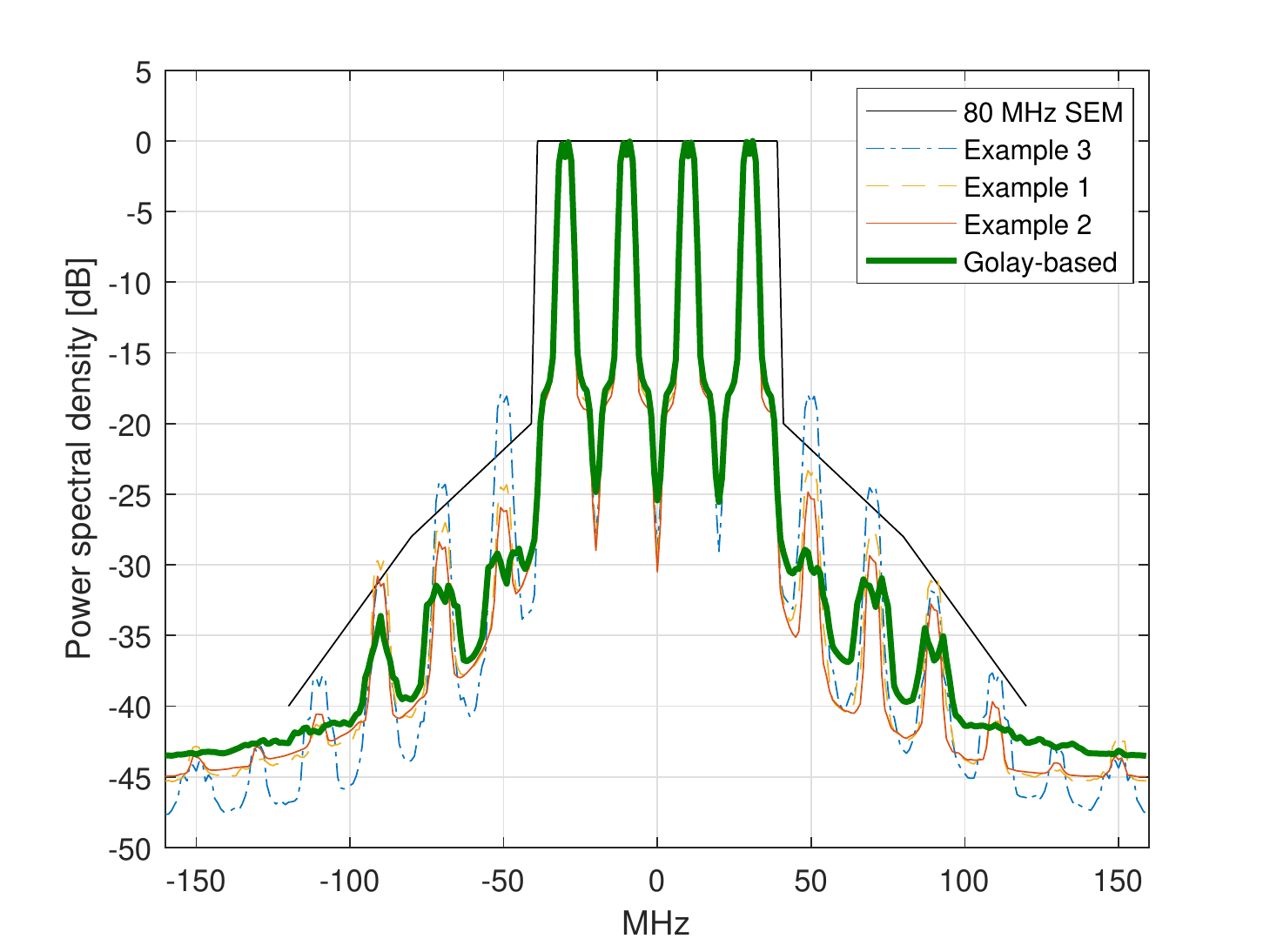}
	}
	\caption{Spectral regrowth (OBO = 5 dB).}
	\label{fig:oob}
	\vspace{-2mm}
\end{figure}

In \figurename~\ref{fig:oob}, we analyze the spectral regrowth under the distortion due to the \ac{PA} non-linearity. For this analysis, we consider a pessimistic scenario where an \ac{AP} wakes up the far stations. We assume that all the channels carry \ac{LDR} \acp{WUS}. We set the \ac{OBO} as $5$ dB. Hence, the P802.11ba packet experiences a heavy distortion. Under this stress test, Example 3 does not satisfy the 802.11 80 MHz \ac{SEM} while Example 2 and Example 1 do not provide room for the spectral regrowth. On the other hand, the proposed method still provides a notable margin under heavy distortion, which indicates robustness against distortion due to the \ac{PA} non-linearity. Hence, the proposed scheme can yield a higher coverage as compared to that of other options under practical scenarios.

\subsection{BER Results}
\begin{figure}[t]
	\centering
	{\includegraphics[width =3.4in]{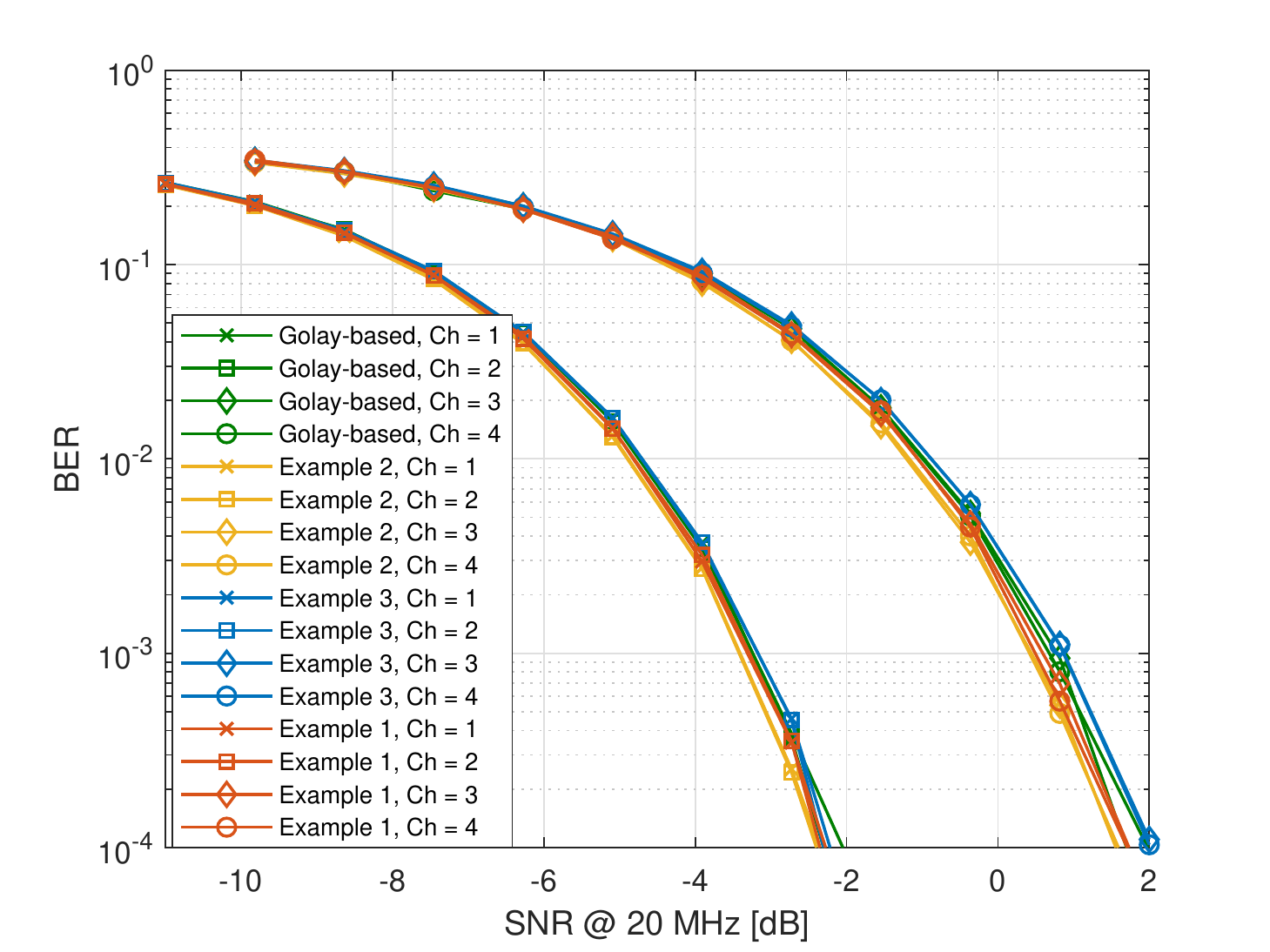}
	}
	\caption{AWGN BER performance without PA.}
	\label{fig:bernoPA}
	\vspace{-2mm}
\end{figure}

\begin{figure}[t]
	\centering
	{\includegraphics[width =3.4in]{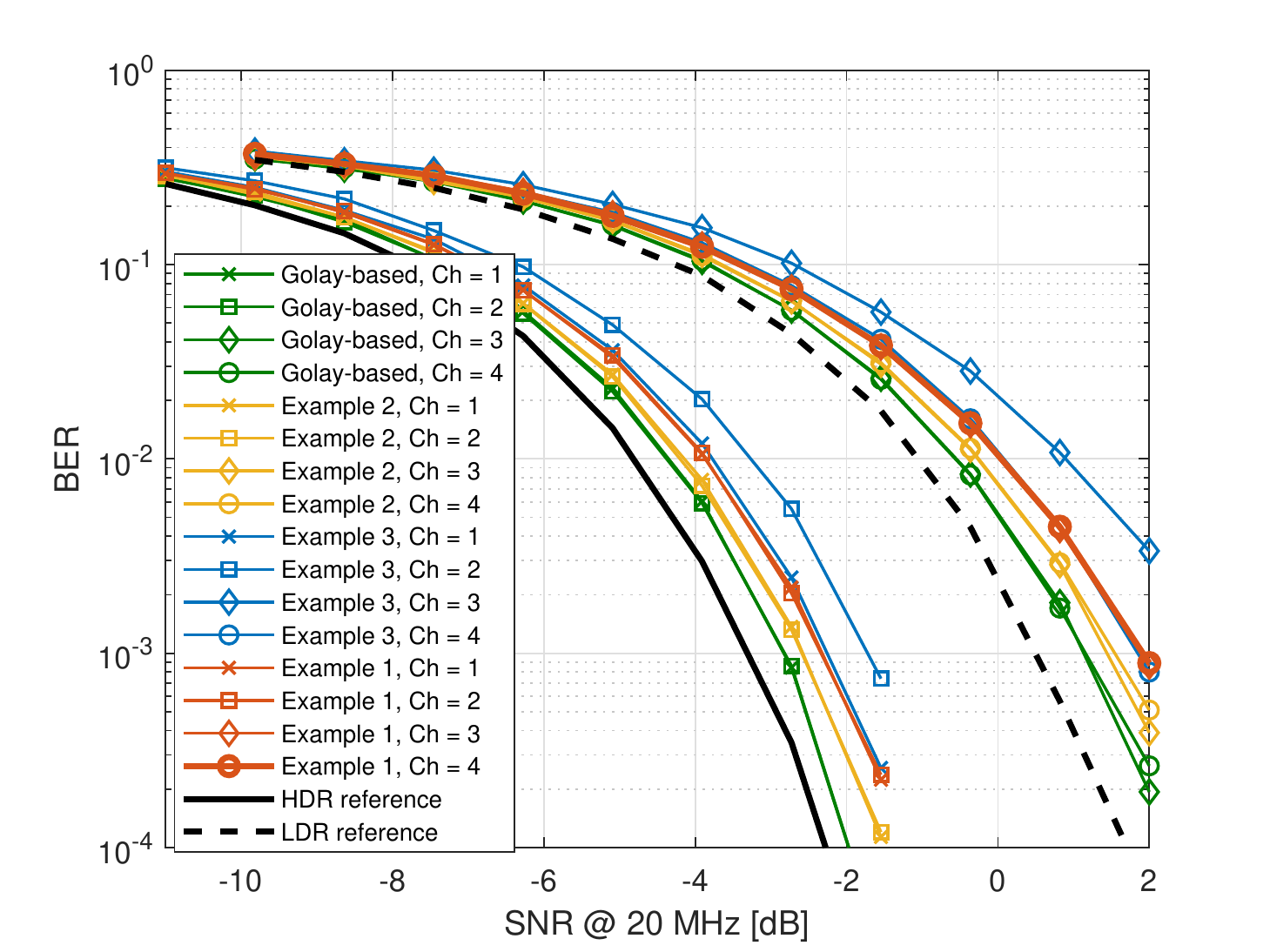}
	}
	\caption{ BER  under AWGN channel and PA impairment (OBO = 5).}
	\label{fig:berPA}
	\vspace{-2mm}
\end{figure}

\begin{figure}[t]
	\centering
	{\includegraphics[width =3.4in]{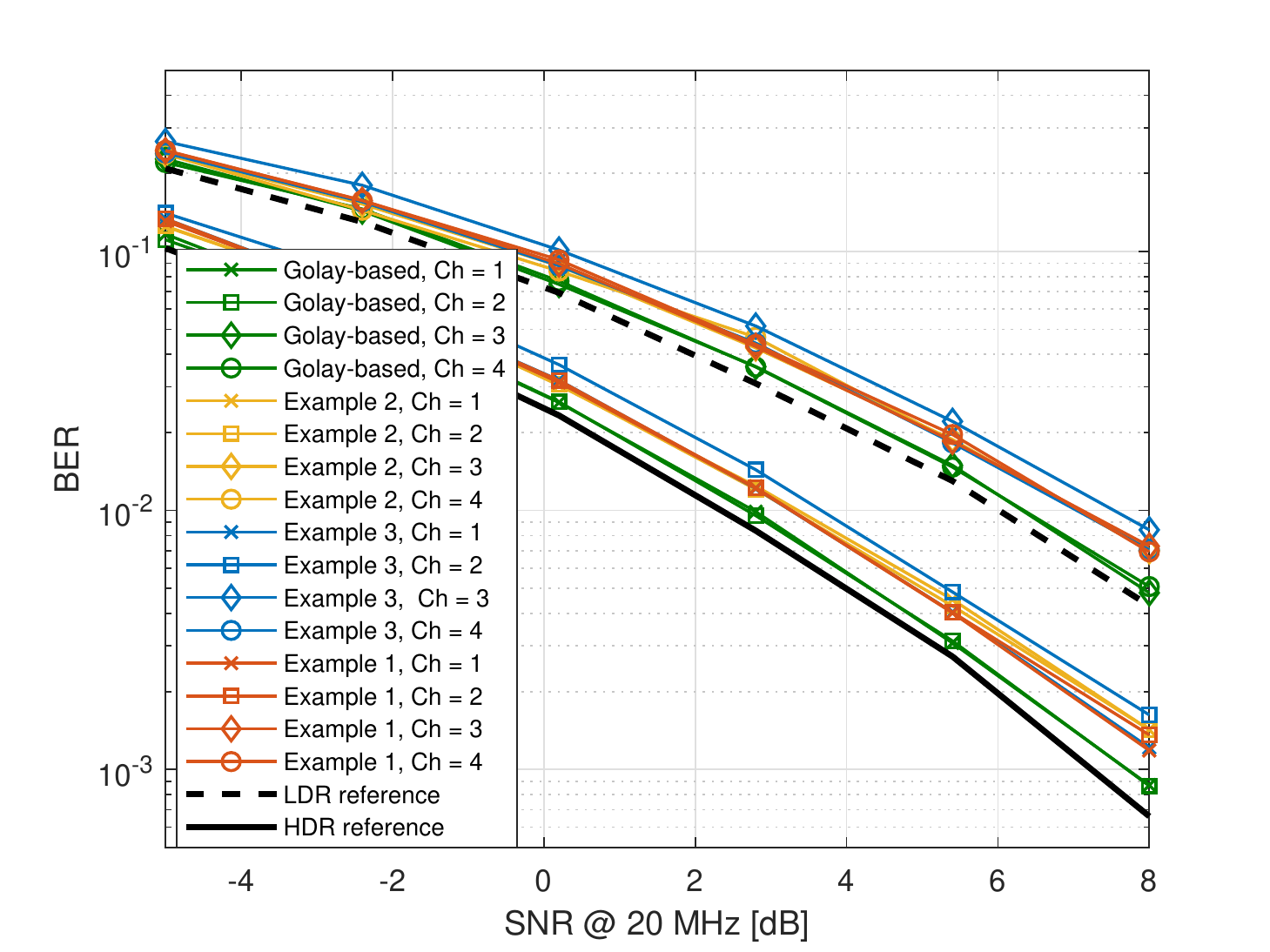}
	}
	\caption{BER  under fading channel  and PA impairment (OBO = 5).}
	\label{fig:berPAf}
	\vspace{-2mm}
\end{figure}

In \figurename~\ref{fig:bernoPA} and \figurename~\ref{fig:berPA}, we provide the \ac{BER} results for an \ac{AWGN} channel. For the configuration, we consider two \ac{HDR} \acp{WUS} and two \ac{LDR} \acp{WUS} are multiplexed in frequency. Without taking the distortion due to the \ac{PA} into account, the \ac{OOK} waveform with the lowest \ac{PAPR} for single channel, i.e., Example 2,  performs slightly better than the others as in \figurename~\ref{fig:bernoPA}. This is because that \ac{WURx} accumulates the sum of the absolute value of the received samples. Nevertheless, the difference among the options is negligible. When heavy \ac{PA} distortion is considered, the \ac{PAPR} of the overall signal becomes more important than the fluctuations within one channel. The \ac{OOK} waveform with better \ac{PAPR} also exhibits a better \ac{BER} performance as in \figurename~\ref{fig:berPA}. While the degradation due to the distortion is negligible for the proposed method for both \ac{LDR} and \ac{HDR} \acp{WUS}, $0.5-1.5$~dB  degradation  in \ac{BER} is observed for the other options, particularly Example 3. The similar results are also obtained under a fading channel as illustrated in \figurename~\ref{fig:berPAf}, where the fading channel model is HyperLAN-A. Note that the \ac{LDR} and \ac{HDR} reference curves  are based on the result for the proposed method  without \ac{PA} impairments and included in \figurename~\ref{fig:berPA} and \figurename~\ref{fig:berPAf} for the sake of comparison. 


\section{Concluding Remarks}\label{sec:conclusion}
The \ac{PAPR} can be large when \acp{WUS} are multiplexed in the frequency domain. Without taking any precautions, the instantaneous peak power can reach 10-12 dB higher than the average power, which can limit the coverage range for \ac{WUS}. The \ac{PAPR} minimization for multiplexed \acp{WUS} is a challenging problem due to the large number of possible cases  in an \ac{FDMA} scenario for IEEE P802.11ba and the simple phase rotations in IEEE 802.11ac cannot address all cases. We show that the existing \acp{CS} can address the \ac{PAPR} issue in the majority of the cases as the corresponding signals are always bounded by 3 dB and the PAPR gain can reach more than $3$~dB in some cases by using known \acp{CS} as compared to the example sequences in \cite{baDraft_2019}.

Although our evaluation shows notable improvement in terms of \ac{PAPR} for P802.11ba FDMA, which has only four channels, how to construct multi-channel \ac{OOK} waveform for any given number of channels still needs to be investigated. In addition, we observe that there may be cases (e.g., when the active channels are asymmetrically located in frequency) where there are no \acp{CS} with \ac{QPSK} alphabet. Hence, another interesting investigation can be an extension to these asymmetric cases to complete the proposed scheme.

\bibliographystyle{IEEEtran}
\bibliography{ookFDMA}

\begin{thebibliography}{10}
\providecommand{\url}[1]{#1}
\csname url@samestyle\endcsname
\providecommand{\newblock}{\relax}
\providecommand{\bibinfo}[2]{#2}
\providecommand{\BIBentrySTDinterwordspacing}{\spaceskip=0pt\relax}
\providecommand{\BIBentryALTinterwordstretchfactor}{4}
\providecommand{\BIBentryALTinterwordspacing}{\spaceskip=\fontdimen2\font plus
\BIBentryALTinterwordstretchfactor\fontdimen3\font minus
  \fontdimen4\font\relax}
\providecommand{\BIBforeignlanguage}[2]{{%
\expandafter\ifx\csname l@#1\endcsname\relax
\typeout{** WARNING: IEEEtran.bst: No hyphenation pattern has been}%
\typeout{** loaded for the language `#1'. Using the pattern for}%
\typeout{** the default language instead.}%
\else
\language=\csname l@#1\endcsname
\fi
#2}}
\providecommand{\BIBdecl}{\relax}
\BIBdecl

\bibitem{McCormick_2017}
D.~K. McCormick, ``{IEEE} technology report on wake-up radio: An application,
  market, and technology impact analysis of low-power/low-latency 802.11
  wireless {LAN} interfaces,'' pp. 1--56, Nov 2017.

\bibitem{baDraft_2019}
``{Part} 11: Wireless {LAN} medium access control ({MAC}) and physical layer
  ({PHY}) specifications amendment: Wake-up radio operation, {Amendment} 9:
  Wake-up radio operation,'' IEEE P802.11ba/D2.0, 2019.

\bibitem{zhang_2018}
H.~Zhang, C.~Li, S.~Chen, X.~Tan, N.~Yan, and H.~Min, ``A low-power
  {OFDM}-based wake-up mechanism for {IoE} applications,'' \emph{IEEE Trans.
  Circuits Syst. II, Express Briefs}, vol.~65, no.~2, pp. 181--185, Feb 2018.

\bibitem{sahin_2018seq}
A.~Sahin and R.~Yang, ``Sequence-based {OOK} for orthogonal multiplexing of
  wake-up radio signals and {OFDM} waveforms,'' in \emph{Proc. IEEE Global
  Telecommunications Conf. (GLOBECOM)}, Dec 2018, pp. 1--5.

\bibitem{wilhelmsson2018}
L.~R. {Wilhelmsson}, M.~M. {Lopez}, S.~{Mattisson}, and T.~{Nilsson},
  ``Spectrum efficient support of wake-up receivers by using {(O)FDMA},'' in
  \emph{Proc. IEEE Wireless Communications and Networking Conference (WCNC)},
  April 2018, pp. 1--6.

\bibitem{sundman_2018}
D.~{Sundman}, M.~M. {Lopez}, and L.~R. {Wilhelmsson}, ``Partial on-off keying -
  a simple means to further improve {IoT} performance,'' in \emph{Proc. IEEE
  Global Internet of Things Summit (GIoTS)}, June 2018, pp. 1--5.

\bibitem{kim_2016}
H.~S. Kim and D.~D. Wentzloff, ``Back-channel wireless communication embedded
  in {WiFi}-compliant {OFDM} packets,'' \emph{IEEE J. Sel. Areas Commun.},
  vol.~34, no.~12, pp. 3181--3194, Dec. 2016.

\bibitem{tang_2015}
S.~Tang, H.~Yomo, and Y.~Takeuchi, ``Optimization of frame length
  modulation-based wake-up control for green {WLAN}s,'' \emph{IEEE Trans. Veh.
  Technol.}, vol.~64, no.~2, pp. 768--780, Feb 2015.

\bibitem{Roberts_2016}
{N. E. Roberts \it et al.}, ``A 236nw -56.5{dBm}-sensitivity {Bluetooth}
  low-energy wakeup receiver with energy harvesting in 65nm {CMOS},'' in
  \emph{Proc. IEEE Int. Solid-State Circuits Conf.}, Jan 2016, pp. 450--451.

\bibitem{Piyare_2017}
{R. Piyare \it et al.}, ``Ultra low power wake-up radios: A hardware and
  networking survey,'' \emph{IEEE Commun. Surveys Tut.}, vol.~19, no.~4, pp.
  2117--2157, Fourth quarter 2017.

\bibitem{steveBin_2018}
S.~Shellhammer, B.~Tian, and R.~van Nee, ``{WUR} power spectral density,'' IEEE
  802.11-18/0824, May 2018.

\bibitem{jiajia_2018}
{J. Jia Jia et. al.}, ``{WUR} preamble sequence design and performance
  evaluation,'' IEEE 802.11-18/0435, 2018.

\bibitem{ieee_2016}
``{Part 11}: Wireless {LAN} medium access control {(MAC)} and physical layer
  {(PHY)} specifications,'' \emph{{IEEE Std 802.11-2016 (Revision of IEEE Std
  802.11-2012)}}, pp. 1--3534, Dec. 2016.

\bibitem{Golay_1961}
M.~Golay, ``Complementary series,'' \emph{IRE Trans. Inf. Theory}, vol.~7,
  no.~2, pp. 82--87, Apr. 1961.

\bibitem{parker_2003}
M.~G. Parker, K.~G. Paterson, and C.~Tellambura, ``Golay complementary
  sequences,'' in \emph{Wiley Encyclopedia of Telecommunications}, 2003.

\bibitem{sahinRui2019ICC}
A.~Sahin and R.~Yang, ``A reliable uplink control channel design with
  complementary sequences,'' in \emph{Proc. IEEE International Conference on
  Communications (ICC)}, May 2019.

\bibitem{holzmann_1991}
W.~H. Holzmann and H.~Kharaghani, ``A computer search for complex {Golay}
  sequences,'' \emph{Aust. Journ. Comb.}, vol.~10, pp. 251--258, Apr. 1994.

\end{thebibliography}

\end{document}